\documentclass[12pt]{iopart}
\usepackage{graphicx}

\begin{document}

\title{The superconductor K$_x$Sr$_{1-x}$Fe$_2$As$_2$: Normal state and superconducting properties}

\author{B. Lv$^1$, M. Gooch$^2$, B. Lorenz$^2$, F. Chen$^2$, A. M. Guloy$^1$ and C. W. Chu$^{2,3,4}$}

\address{$^1$ TCSUH and Department of Chemistry, University of Houston, Houston, TX 77204, USA}
\address{$^2$ TCSUH and Department of Physics, University of Houston, Houston, TX 77204, USA}
\address{$^3$ Lawrence Berkeley National Laboratory, 1 Cyclotron Road, Berkeley, CA 94720, USA}
\address{$^4$ Hong Kong University of Science and Technology, Hong Kong, China}

\begin{abstract}
The normal state and superconducting properties are investigated in the phase diagram of K$_x$Sr$_{1-x}$Fe$_2$As$_2$ for 0$\leq$x$\leq$1. The
ground state upper critical field, H$_{c2}$(0), is extrapolated from magnetic field dependent resistivity measurements. H$_{c2}$(0) scales with
the critical temperature, T$_c$, of the superconducting transition. In the normal state the Seebeck coefficient is shown to experience a
dramatic change near a critical substitution of x$\simeq$0.3. This is associated with the formation of a spin density wave state above the
superconducting transition temperature. The results provide strong evidence for the reconstruction of the Fermi surface with the onset of
magnetic order.
\end{abstract}

\maketitle

\section{Introduction}
The discovery of high-temperature superconductivity in rare earth (R) oxypnictides, ROFeAs \cite{kamihara:08}, has initiated extensive research
activities in this class of compounds \cite{chen:08,chen:08b,takahashi:08,ren:08,ren:08b}. The main structural building block of the FeAs-based
superconductors is the Fe$_2$As$_2$ layer with tetrahedrally coordinated Fe covalently bonded to 4 As atoms. The Fe$_2$As$_2$ blocks are the
active layers for the observed superconductivity and they are separated by charge reservoir layers that can modify or change the average charge
of the active layer through appropriate chemical substitutions. In ROFeAs the appropriate doping can be achieved by substituting F for O
\cite{chen:08c,kohama:08,lee:08,luetkens:08,zhao:08} or by removing oxygen \cite{ren:08c}. However, the amount of charge carriers introduced is
limited and attempts to achieve higher carrier densities in the Fe$_2$As$_2$ layers of the ROFeAs system frequently results in chemical
disproportionation and decomposition and the formation of impurity phases \cite{kamihara:08,chen:08,chen:08c,kohama:08,lee:08,zhao:08}.

The Fe$_2$As$_2$ layers are very rigid and the charge reservoir blocks can be replaced by different oxygen-free layers, for example, by single
layers of alkali metal or alkaline earth (Ae=Ca, Sr, Ba) ions. Stable structures of KFe$_2$As$_2$, CsFe$_2$As$_2$ \cite{sasmal:08}, and LiFeAs
\cite{wang:08,pitcher:08,tapp:08} have been synthesized and found superconducting at different temperatures. It is interesting that the alkali
metal-based compounds apparently do not need any additional doping to induce superconductivity, and they belong to a small family of intrinsic
(self-doped) FeAs superconductors. The critical temperatures of KFe$_2$As$_2$ and CsFe$_2$As$_2$ are relatively low, below 4 K, but can be
significantly increased up to 37 K by replacing the alkali metals with alkaline earth metals (corresponding to a change of the carrier density
in the Fe$_2$As$_2$ layer) \cite{sasmal:08,rotter:08}. The AeFe$_2$As$_2$ compounds are nonsuperconducting but they show a transition into a
spin density wave (SDW) state at higher temperatures. The continuous substitution of K for Sr or Ba in K$_x$(Sr,Ba)$_{1-x}$Fe$_2$As$_2$ leads to
the onset of superconductivity at a critical doping of x$_c$=0.17, an increase of T$_c$ to a maximum T$_c\simeq$37 K near the optimal doping
(x$_{opt}\simeq$0.45) and, with further increasing x, a decrease of T$_c$ to 3.8 K at x=1 (KFe$_2$As$_2$). Thus a complete superconducting phase
diagram is obtained at all doping levels between x=0 and x=1 \cite{sasmal:08,rotter:08b,chen:08d}. The doping dependence of T$_c$ in this system
seems to be qualitatively similar to the high-T$_c$ copper oxide superconductors suggesting that the underlying physics in both systems is
determined by the control of the charge density in the active Fe$_2$As$_2$ layers. This conjecture recently received support from a
high-pressure investigation of K$_x$Sr$_{1-x}$Fe$_2$As$_2$ \cite{gooch:08}. One major difference of the FeAs- and CuO-based high-temperature
superconductors is the existence or coexistence of magnetic order in the carrier density-dependent phase diagram. In the copper oxides
antiferromagnetism exists in the undoped, insulating state and it is completely suppressed before superconductivity arises with doping. In
K$_x$Sr$_{1-x}$Fe$_2$As$_2$ magnetic order in form of a spin density wave has been reported for the non-superconducting SrFe$_2$As$_2$
\cite{krellner:08} and BaFe$_2$As$_2$ \cite{rotter:08c}. The SDW transition is suppressed with doping of potassium, however, in a narrow range
of the phase diagram, for 0.17$<$x$<$0.3, the spin density wave and superconducting transitions appear as successive transitions upon cooling
\cite{chen:08d,gooch:08}. This raises the question about the coexistence of magnetic and superconducting orders in different regions of the
phase diagram and it demands a detailed investigation of the physical properties in the whole phase diagram of K$_x$(Sr,Ba)$_{1-x}$Fe$_2$As$_2$
with special attention to the narrow region between 0.17 and 0.3.

We have therefore synthesized K$_x$Sr$_{1-x}$Fe$_2$As$_2$, for 0$\leq$x$\leq$1, and measured the normal state and superconducting transport
properties. We show that the superconducting phases exhibit very high upper critical fields that scale with the transition temperatures. The
Seebeck coefficients, S(T), of the series of samples are positive over the whole temperature range (except for x$\simeq$0), and S exhibits a
significant change at x$\simeq$0.3 where the spin density wave state forms at higher temperatures.

\section{Synthesis and Experimental}
Polycrystalline samples of K$_x$Sr$_{1-x}$Fe$_2$As$_2$ were prepared from stoichiometric amounts of the ternary end members, KFe$_2$As$_2$ and
SrFe$_2$As$_2$, thoroughly mixed, compressed, and annealed in welded Nb containers (jacketed in quartz) at 700 $^\circ$C for 100 to 120 hours.
The high phase purity of the ternary compounds, prepared from high purity K, Sr, and FeAs as described previously \cite{sasmal:08}, guarantees
the exceptional quality of the solid solutions and the precise control of the chemical compositions. Electrical conductivities were measured by
a four probe technique using the low-frequency (19 Hz) resistance bridge LR700 (Linear Research) and a Physical Property Measurement System
(Quantum Design) for temperature (T$>$ 1.8 K) and magnetic field (H$<$ 7 Tesla) controls. The Seebeck coefficients were measured between 300 K
and 5 K employing a high precision ac technique with a dual heater system \cite{choi:07}. The error in the measurements is about 0.05 $\mu$V/K.

\section{Results and Discussion}
\subsection{Structure and lattice parameters}
The room temperature X-ray diffraction data of all samples were indexed to the ThCr$_2$Si$_2$-type tetragonal I4/mmm structure. The systematic
change of the a- and c-axis lattice constants (Fig. 1) indicate the formation of a uniform solution between the two compounds KFe$_2$As$_2$ and
SrFe$_2$As$_2$. The increase of the c-axis by nearly 12 \% with increasing x is consistent with the larger K atom replacing the smaller Sr atom.
The change of the a-axis (2 \% decrease from SrFe$_2$As$_2$ to KFe$_2$As$_2$) is much smaller. The systematic changes in the unit cell
parameters result in an increase of the volume as well as the c/a ratio with x.

\begin{figure}
\begin{center}
\includegraphics[angle=0, width=3.5 in]{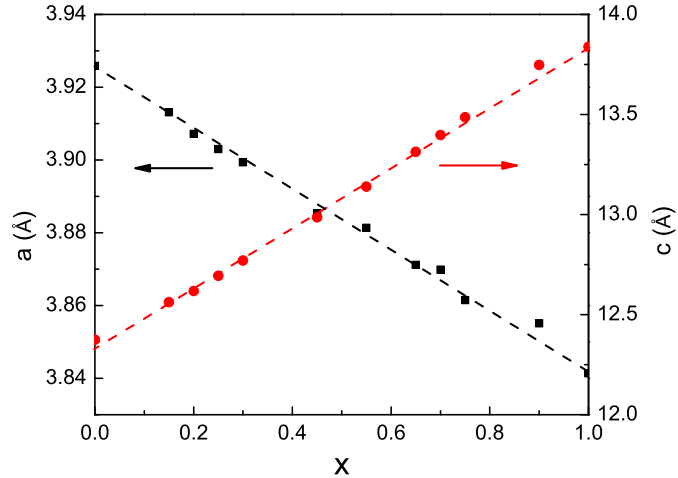}
\end{center}
\caption{Lattice parameters of K$_{x}$Sr$_{1-x}$Fe$_2$As$_2$ vs. composition x.}
\end{figure}

\subsection{Resistivity at zero field: The phase diagram}
The resistivity of SrFe$_2$As$_2$ shows the distinct anomaly (inflection point) at the spin density wave (SDW) transition temperature, T$_S$=204
K, as reported earlier \cite{krellner:08}. This anomaly is very sharp and it was shown recently that the SDW transition in SrFe$_2$As$_2$
coincides with a structural deformation accompanied by a change of symmetry from tetragonal (T$>$T$_S$) to orthorhombic (T$<$T$_S$)
\cite{kaneko:08,jesche:08}. With increasing K content, the SDW anomaly shifts to lower temperatures as shown in Fig. 2. For x=0.12, the SDW
transition temperature decreased to 175 K from 204 K (x=0). With further increasing x, the critical temperature T$_S$ drops quickly and the
sharp anomaly of dR/dT disappears near x$\simeq$0.3, limiting the SDW state to the very strontium-rich region of the phase diagram. It is
interesting to note that, before the SDW state is completely suppressed, superconductivity arises at a K concentration of about 0.17, with a
T$_c$ of 2.9 K (Fig. 2). Therefore, in a narrow range of chemical composition, between x=0.17 and x=0.3, K$_x$Sr$_{1-x}$Fe$_2$As$_2$ passes from
a nonmagnetic into a SDW state at T$_S$ and then into the superconducting state at T$_c$. It is not clear, however, if the SDW order survives
the transition into the superconducting state, and if both orders coexist at low temperatures.

\begin{figure}
\begin{center}
\includegraphics[angle=0, width=4 in]{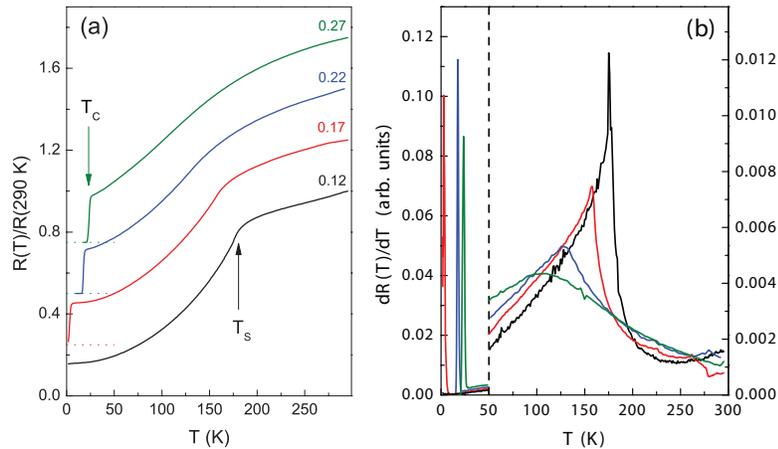}
\end{center}
\caption{Resistivity (a) and its derivative (b) of K$_{x}$Sr$_{1-x}$Fe$_2$As$_2$ for x$<$0.3. T$_S$ and T$_c$ are defined by the sharp peaks of
dR/dT in (b). The curves in (a) are vertically offset for better clarity and the attached numbers are the values of x. Note that the vertical
scale in (b) is changed at 50 K.}
\end{figure}

Upon further increasing x (hole doping) the superconducting T$_c$ rises to a maximum of 37 K and decreases again between x$_{max}$=0.45 and x=0.
The resistivity data for x$>$0.3 are shown in Fig. 3. The temperature derivative, dR/dT, clearly defines the critical temperature and shows the
sharp resistive transition into the superconducting state. The phase diagram, derived from the resistivity data and shown in Fig. 4, is slightly
asymmetric with respect to x. The nearly parabolic T$_c$(x) curve develops a tail-like extension to large values of x, and the ternary
KFe$_2$As$_2$ remains superconducting with a T$_c$ of 3.8 K. The major part of the phase diagram is similar to that known for the high-T$_c$
cuprate superconductors having a dome-shaped parabolic T$_c$ dependence on the carrier density. The main qualitative difference is the tail of
the phase boundary in the K$_x$Sr$_{1-x}$Fe$_2$As$_2$ system.

\begin{figure}
\begin{center}
\includegraphics[angle=0, width=2.5 in]{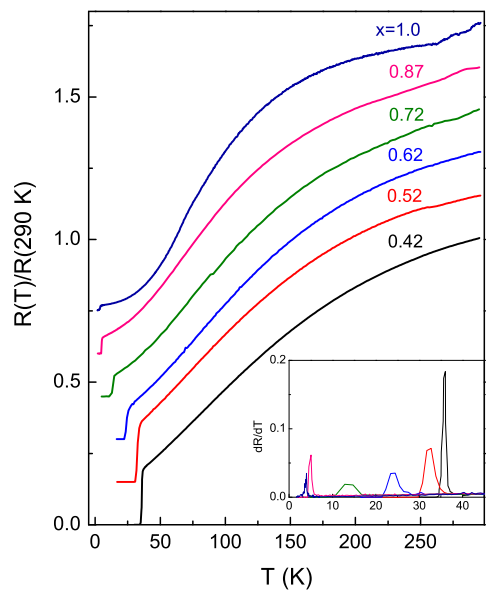}
\end{center}
\caption{Resistivity of K$_{1-x}$Sr$_{x}$Fe$_2$As$_2$ for x$<$0.7. Different curves are vertically offset and separated by 0.15 units to enhance
clarity of presentation. Inset: Derivative dR/dT with the peak position defining T$_c$.}
\end{figure}

\begin{figure}
\begin{center}
\includegraphics[angle=0, width=3.5 in]{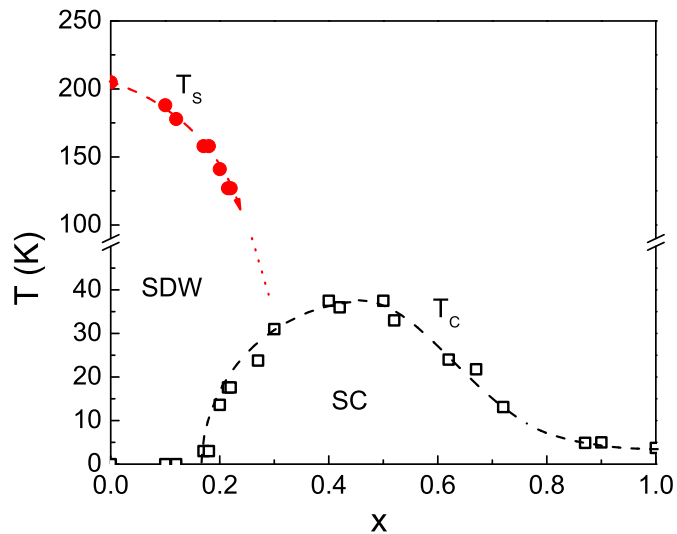}
\end{center}
\caption{T-x phase diagram of K$_{x}$Sr$_{1-x}$Fe$_2$As$_2$. The dashed lines serve as a guide to the eye.}
\end{figure}

The extrapolation of the SDW phase boundary to lower temperatures suggests that the magnetically ordered state becomes unstable for values of
x$>$0.3. The coexistence of both, the SDW and the superconducting transitions between x=0.17 and x=0.3 is another distinct feature that is
different from the cuprate superconductors. In the latter system the antiferromagnetic order of the copper spins is completely suppressed before
superconductivity sets in at a higher doping level, and both transitions do not coexist at any point of the T$_c$(n) (n carrier density) phase
diagram. The extrapolation of T$_S$ to zero temperature, as schematically shown in Fig. 4 (dotted line), may suggest a "hidden" quantum critical
point (QCP) at T=0 and superconductivity induced by magnetic fluctuations. The exact value of x defining the QCP is difficult to determine since
the SDW phase boundary is cut off by the superconducting transition and, as shown in Fig. 2 (b), the resistive anomaly that is characteristic
for the SDW transition broadens significantly with increasing x. If magnetic fluctuations are involved in the superconducting pairing mechanism
one should expect a maximum of T$_c$ close to the critical doping x of the QCP, where the magnetically ordered state comes closest to the
superconducting state. In the phase diagram of Fig. 4, the highest T$_c$ is observed at a K content of about 0.45. More detailed experiments are
therefore needed near the maximum composition of the superconducting dome. Alternatively, the SDW and superconducting states may compete for the
ground state in which case magnetic order would be counterproductive in stabilizing the superconducting state. Recent results of high-pressure
experiments have shown that, in the narrow coexistence range of both transitions, the application of external pressure did drastically reduce
the SDW temperature whereas the superconducting T$_c$ was enhanced \cite{gooch:08}, suggesting a possible competition between both states.

\subsection{Normal state properties: The Seebeck coefficient}
The electrical resistivity of polycrystalline samples depends, to a large extent, on the grain boundary resistance and contributions from
grain-grain contacts. It is therefore difficult to extract detailed information about the electronic states. The Seebeck coefficient is less
sensitive to grain boundary effects (since it is a zero-current property) and it provides information about both thermodynamic and transport
properties of the carriers. Its sensitivity to the electronic band structure makes it a perfect probe in detecting changes in the density of
states and the Fermi surface. The Seebeck coefficient for a single band metal can be expressed as \cite{mott:62}
\begin{equation}
S=\frac{\pi^2k^{2}_{B}T}{3e}\left[\frac{d\ln D(E)}{dE}+\frac{d\ln \nu^2(E)}{dE}+\frac{d\ln \tau(E)}{dE}\right]
\end{equation}
D(E), $\nu$(E), and $\tau$(E) are the density of states, an average charge velocity, and the carrier scattering relaxation time, respectively.
The first term in the brackets shows the sensitivity of S with respect to changes of the band structure, shifts of the Fermi energy, or
instabilities of the Fermi surface.

\begin{figure}
\begin{center}
\includegraphics[angle=0, width=3.5 in]{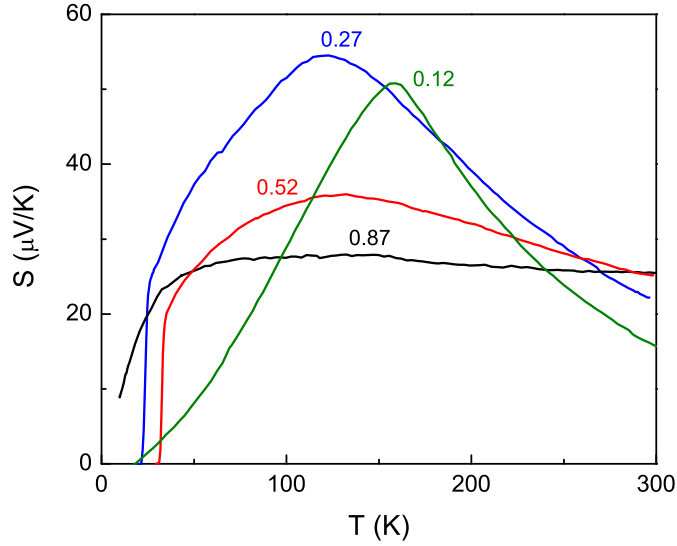}
\end{center}
\caption{Seebeck coefficient of K$_x$Sr$_{1-x}$Fe$_2$As$_2$ for selected values of x (labels next to the curves).}
\end{figure}

The Seebeck coefficients, S(T), for some selected compositions of K$_x$Sr$_{1-x}$Fe$_2$As$_2$ are shown in Fig. 5. S is positive over the whole
temperature range (with the exception of SrFe$_2$As$_2$ where it is negative near room temperature and at low T), and drops to zero at the
superconducting transition. The positive sign is consistent with the positive Hall coefficient \cite{chen:08e} indicating that the majority
carriers are holes in the system K$_x$Sr$_{1-x}$Fe$_2$As$_2$. However, band structure calculations \cite{singh:08} and ARPES measurements
\cite{liu:08b} have shown that the Fermi surface consists of states from four different bands, some with electron-like and others with hole-like
states. The hole Fermi surfaces are located around the zone center whereas the electron surfaces are at the zone corner (M point) of the
primitive tetragonal Brillouin zone. Contributions from all bands to the Seebeck coefficient are therefore expected and the doping and
temperature dependence of S(T) will be more complex than expressed by the simple formula (1).

The room temperature values of S initially increase with x as a consequence of hole doping. This could be explained within a rigid band model by
a lowering of the Fermi energy and an increase of the density of states. However, results of a recent density functional band structure
calculation within a virtual crystal approximation \cite{singh:08} have questioned the validity of the rigid band model and have shown (for the
related system K$_x$Ba$_{1-x}$Fe$_2$As$_2$) that the density of states depends only weakly on the doping level. The major effect of hole doping
is a redistribution of states between the electron and hole pockets of the Fermi surface without significant change of the total N(E$_F$). With
increasing x the hole states tend to dominate as is also obvious from the positive signs of the Seebeck coefficient and the Hall number. This
redistribution of states with x explains the increase of the room temperature value of S. Since different contributions from electron and hole
carriers to S(T) partially cancel out, the weight shift towards the hole pockets at the Fermi surface results in an increase of S(T) even if
N(E$_F$) is not changed. The effect is even larger at lower temperatures and it is also reflected in the T-dependence of S.

For small values of x, S(T) shows a pronounced maximum near the SDW transition. The maximum value, S$_{max}$, rapidly increases with hole doping
until x reaches a critical value x$\simeq$0.3. At the critical x, the value of S$_{max}$ peaks and decreases with further doping. The
x-dependence of S$_{max}$ is shown in Fig. 6. The sharp and distinct peak at the critical composition indicates a dramatic change of the Fermi
surface and the density of states with the onset of the SDW order. This might be an indication of an important change in the electronic system
and the possible existence of a magnetic quantum phase transition at T=0 that could only be revealed if the superconductivity was completely
suppressed. Theoretical calculations show that the SDW state is close in energy to other magnetically ordered states such as ferromagnetism and
checkerboard antiferromagnetism. Whenever different magnetic states are competing for the ground state spin fluctuations in the paramagnetic
phase may be large and significant. The doping-induced instability of the magnetic order in the presence of large spin fluctuations, possibly
giving rise to the pairing interaction in the superconducting state, can then be understood as a delicate balance between the SDW and the
superconducting states. Experiments at much lower temperature are suggested to reveal more details about the critical behavior at the SDW phase
boundary and the QCP. However, as shown in the phase diagram of Fig. 4, the onset of superconductivity at about 30 K (for x$\simeq$0.3) does not
allow the study of the SDW phase boundary closer to zero temperature. The suppression of superconductivity, for example in high magnetic fields,
is therefore suggested to investigate the QCP. However, as discussed in the following section, the superconducting critical field in this
composition range is very high and static fields of this magnitude may not be available. It is also not clear how the high field would affect
the SDW state.

\begin{figure}
\begin{center}
\includegraphics[angle=0, width=3.5 in]{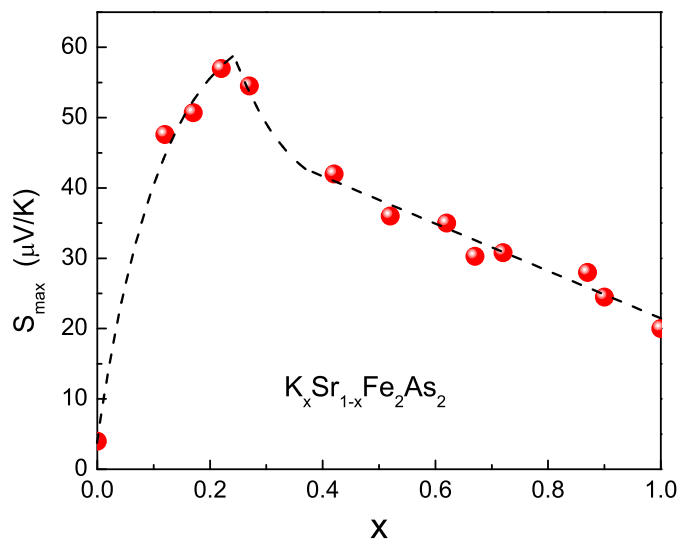}
\end{center}
\caption{The maximum of the Seebeck coefficient as a function of x. The peak of S$_{max}$(x) appears at the onset x$_S$ of the SDW phase.}
\end{figure}

\subsection{Upper critical field of the superconducting state}
The dependence of superconducting properties on external magnetic fields is an essential property which determines the performance of
superconducting devices in magnets, for energy transport, etc. For type-II superconductors, the upper critical field (H$_{c2}$) limits the
stability range of the superconducting state and its zero temperature value, H$_{c2}$(0), is a characteristic quantity that can be determined or
extrapolated in high-field measurements. In the K$_x$Sr$_{1-x}$Fe$_2$As$_2$ system H$_{c2}$(0) can be very high, exceeding 100 Tesla, as shown
for the optimally doped compounds \cite{sasmal:08}.

\begin{figure}
\begin{center}
\includegraphics[angle=0, width=5 in]{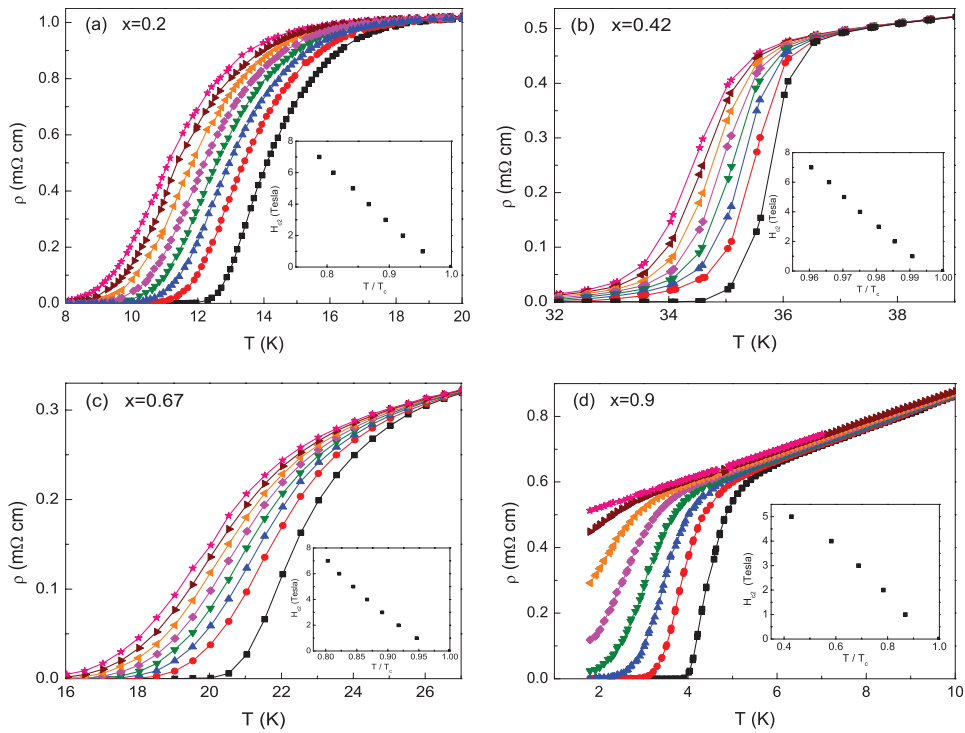}
\end{center}
\caption{Resistivity of K$_x$Sr$_{1-x}$Fe$_2$As$_2$ in magnetic fields at the superconducting transition for different compositions x. Different
curves are measured in magnetic fields from 0 to 7 Tesla in steps of 1 Tesla (right to left, black to pink). The insets show the critical fields
estimated from the midpoint of the resistivity drop.}
\end{figure}

\begin{figure}
\begin{center}
\includegraphics[angle=0, width=3.5 in]{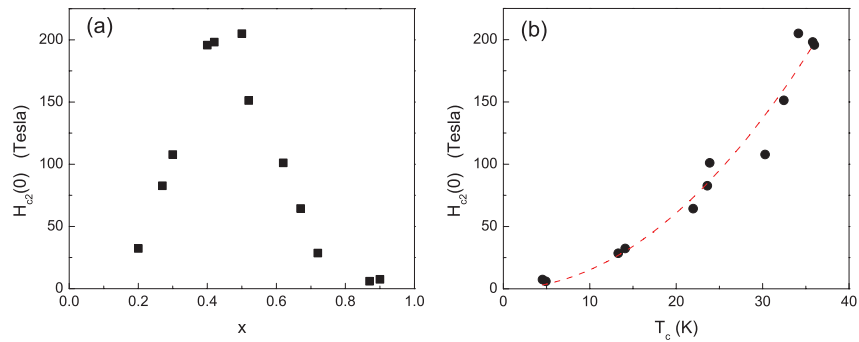}
\end{center}
\caption{The upper critical field, H$_{c2}$(0) as function of x (a) and T$_c$ (b).}
\end{figure}

To extrapolate the zero temperature H$_{c2}$ for the complete series of K$_x$Sr$_{1-x}$Fe$_2$As$_2$ compounds we measured the resistivity in
magnetic fields up to 7 Tesla. Some selected data for x=0.2, 0.42, 0.67, and 0.9 are shown in Figs. 7 (a) to (d). The insets to the figures show
the temperature dependence of H$_{c2}$. T$_c$ at different fields was determined from the midpoint of the resistivity drop. This T$_c$ criteria
may appear somewhat arbitrary and the exact value of the extrapolated H$_{c2}$(0) may depend on the way T$_c$ was defined (alternative criteria,
for example, are the 10 \% or 90 \% resistivity drop), however, since we are more interested in comparing the estimated H$_{c2}$ values for
different x throughout the phase diagram the choice of the 50 \% resistivity drop is reasonable. The zero temperature H$_{c2}$(0) was
extrapolated using the Ginzburg-Landau formula, $H_{c2}(t)=H_{c2}(0)*(1-t^2)/(1+t^2)$, with $t=T/T_c$, from the data similar to those shown in
the insets to Fig. 7. The estimated values are within 5 \% of the H$_{c2}$ determined from the alternative relation $H_{c2}(0)\simeq0.7*T_c*\mid
dH_{c2}(T)/dT\mid_{T_c}$ using the slope of H$_{c2}$(T) near T$_c$.

The extrapolated values for H$_{c2}$(0) as a function of the K content, x, are shown in Fig. 8 (a). H$_{c2}$(0) is very large, up to 200 Tesla,
in the optimally doped region of the phase diagram with T$_c$'s up to 37 K. However, it is drastically reduced, on both sides, where T$_c$ drops
to zero (x$<$x$_{max}$) or to small values for x$\rightarrow$1. This suggests that H$_{c2}$ scales with the critical temperature. Plotting
H$_{c2}$(0) versus T$_c$ in Fig. 8 (b) indeed shows a parabolic dependence, $H_{c2}(0)=A*T_c^2$ (dashed line in Fig. 8 (b), A=0.152
Tesla/K$^2$). The large values of H$_{c2}$ are typical for a short Ginzburg-Landau coherence length, $\xi_{GL}$, that can be extracted through
$H_{c2}(0)=\Phi_0/[2\pi\xi^2_{GL}]$, where $\Phi_0$ is the flux quantum. For the highest T$_c$'s in the K$_x$Sr$_{1-x}$Fe$_2$As$_2$ system
coherence lengths of about 13 {\AA} are thus obtained. The short coherence length in the superconducting state is another indication of the
analogy of the FeAs-based superconductors to the high-T$_c$ copper oxides. While the discussion about the pairing symmetry and the structure of
the superconducting gap in FeAs superconductors is not yet decided there is increasing evidence for the existence of nodes in the
superconducting gap function and two-gap superconductivity with carriers from two bands forming Cooper pairs \cite{ding:08}. A more detailed
investigation of the superconducting properties of the FeAs superconductors has to take into account the specifics of the pairing symmetry and
the different bands at the Fermi surface.

\section{Summary}
We have investigated the phase diagram and the normal state transport properties of the superconducting system K$_x$Sr$_{1-x}$Fe$_2$As$_2$. The
dome-shaped phase diagram of the superconducting state resembles that of the high-T$_c$ cuprates with a maximum of T$_c$ near an optimal doping
range. Magnetic order exists on the Sr-rich side and the spin density wave transition is followed by a superconducting transition within a
narrow doping range. The Seebeck coefficient shows a distinct anomaly near x=0.3 where the SDW phase arises above the superconducting dome. This
anomaly indicates a significant change of the Fermi surface and the possible existence of a quantum critical point is proposed. The most
interesting property of the superconducting state is a very high upper critical field estimated from resistivity measurements in magnetic
fields. The large values of H$_{c2}$(0) imply a short Ginzburg-Landau coherence length, similar to the high-T$_c$ cuprate superconductors.

\ack This work is supported in part by the T.L.L. Temple Foundation, the J.J. and R. Moores Endowment, the State of Texas through TCSUH, the
USAF Office of Scientific Research, and at LBNL through USDOE. A.M.G. and B.Lv acknowledge the support from the NSF (CHE-0616805) and the R.A.
Welch Foundation.

\section*{References}
\bibliographystyle{unsrt}

\end{document}